# Performance Improvement of OFDM System Using Iterative Signal Clipping With Various Window Techniques for PAPR Reduction

Smita Jolania[#1], Sandeep Toshniwal[*2]

[#1]*M.Tech*(Digital Communication)*
*Kautilya Institute of Technology and Engineering,RTU*
[2]*Reader(Electronics and communication)*
*Kautilya Institute of Technology and Engineering,RTU*
*Jaipur,India*

*Abstract*— **OFDM signals demonstrates high fluctuations termed as Peak to Average Power Ratio (PAPR).The problem of OFDM is the frequent occurrence of high Peaks in the time domain signal which in turn reduces the efficiency of transmit high power amplifier.In this paper we discussed clipping and filtering technique which is easy to implement and reduces the amount of PAPR by clipping the peak of the maximum power signal.This technique clips the OFDM signal to a predefined threshold and uses a filter to eliminate the out-of-band radiation.Moreover, analysis of PAPR is given by varying different filters.The study is focused to reduce PAPR by iterative clipping and filtering method. The symbol error rate performances for different modulation techniques have been countered.Each clipping noise sample is multiplied by a window function(e.g.Hanning,Kaiser, or Hamming) to suppress the out-of-band noise.It is shown that clipping and different filtering techniques for improvement in the SER performance and provides further reduction in PAPR.**

## INTRODUCTION

Orthogonal Frequency-Division Multiplexing (OFDM) is one of the technologies considered for 4G broadband wireless communications due to its robustness against multipath fading and relatively simple implementation compared to single carrier systems.OFDM works by splitting the radio signal into multiple smaller sub-signals which are then transmitted simultaneously at different frequencies to the receiver.The basic idea is using a large number of parallel narrow-band subcarriers instead of a single wide-band carrier to transport information.This method is very easy and efficient in dealing with multi-path and robust again narrow-band interference.These sub-carriers (or sub-channels) divide the available bandwidth and are sufficiently separated in frequency (frequency spacing) so that they are orthogonal.The orthogonality of the carriers means that each carrier has an integer number of cycles over a symbol period. Due to this, the spectrum of each carrier has a null at the center frequency of each of the other carriers in the system.This results in no interference between the carriers, although their spectra overlap. The separation between carriers is theoretically minimal so there would be a very compact spectral utilization. OFDM systems are attractive for the way they handle ISI, which is usually introduced by frequency selective multipath fading in a wireless environment.To improve the spectral efficiency of OFDM we eliminate band guards between carriers by using orthogonal carriers (allowing overlapping). But it is limited by the problem of Peak-to-average power efficiency of RF amplifier at the transmitter.

### I. OFDM SYSTEM DESCRIPTION AND PAPR

A Basic OFDM system has an input data symbols are supplied into a channel encoder that data are mapped onto BPSK/QPSK/QAM constellation.The data symbols are converted from serial to parallel and using Inverse Fast Fourier Transform (IFFT) to achieve the time domain OFDM symbols.

For an OFDM system with N subcarriers, OFDM signal in baseband notation in discrete form for interval $mT_u \le t \le (m+1)T_u$ can be expressed as:

Xn = IFFT{Xk}

$$X(t) = \frac{1}{\sqrt{N}} \sum_{k=1}^{N-1} X_k e^{j2\pi k \Delta f t} \qquad (1)$$

where, $f_0 = 1/T$. Replacing $t = n \cdot T_b$ where, $T_b = T/N$
$X_k$ is transmitted symbol on the $K^{th}$ subcarrier
N is number of subcarriers

Mathematically the continuous time representation of the OFDM transmit signal in the discrete time version can be given by:

$$X_m(t) = \frac{1}{\sqrt{N}} \sum_{k=0}^{N-1} X_k e^{2\pi km/N} \qquad (2)$$

The frequency difference between subcarriers=Δf

The PAPR of the signal, x (t) is then given as the ratio of the peak instantaneous power to the average power, written as

$$PAPR = \frac{max_{0 \leq t \leq T} |s(t)|^2}{E\{|s(t)|^2\}} \qquad (3)$$

As more sub-carriers are added, higher peak values may occur, hence the PAPR increases proportionally with the number of sub-carriers.At the transmitter side, the invertible clipping method reduces the amplitude dynamics and thus the PAPR of the signal that has to be amplified. This result is presented in Complementary Cumulative Distribution Function (CCDF).

## II. CLIPPING AND FILTERING

In the clipping technique hard limiting is applied to the amplitude of the complex values of the IFFT output.The filtering technique is designed to alleviate out-of-band distortion but cannot correct in-band distortion.The filtering operation will lead to peak power regrowth.By repeating clipping several times, we can reduce the likelihood of peak power regrowth.Such a procedure is called (Recursive Clipping and filtering)RCF.

A. Firstly the data signal X is transformed into the time domain signal using an IFFT.After an IFFT, the original signal is clipped in the time domain. The clipping can be described by the equation below:

$$C = \begin{cases} |x|e^{j\phi}, & |x| \leq A \\ Ae^{j\phi}, & |x| > A \end{cases} \qquad (4)$$

where C represents the output of the time domain signal, A is the threshold clipping level

B. The clipped time domain signal C is then converted back into the frequency domain using an FFT.Filtering is then performed to remove out of band power.

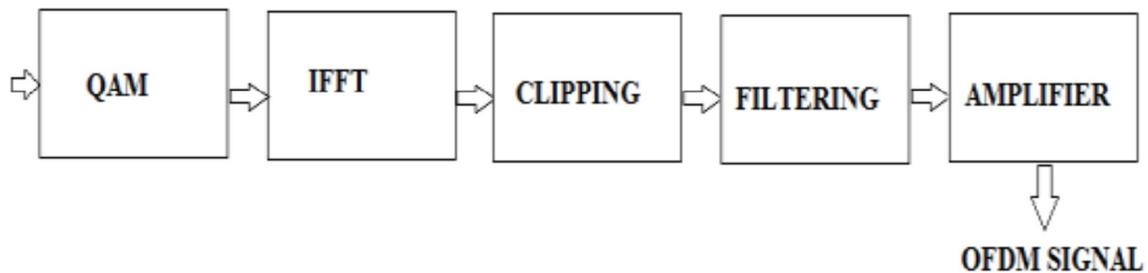

Figure 1. OFDM signal transmission block diagram

## IV. SIMULATION OF PROPOSED METHOD

In this section, the PAPR performance is evaluated by simulations in MATLAB. In this simulation QAM symbols are fed as input to the OFDM system.In proposed clipping technique using single iteration PAPR is reduced up to small value compared to unclipped signal.Further it is shown in figure that when number of iterations are increased to five PAPR is further reduced.PAPR further can go down by applying filters.

The algorithm for clipping is as follows:
1) Convert the OFDM symbol to time domain as IFFT.
2) Clip to the threshold
3) The clipped OFDM signal then filtered using different filters.
4) Convert to time domain and transmit it.

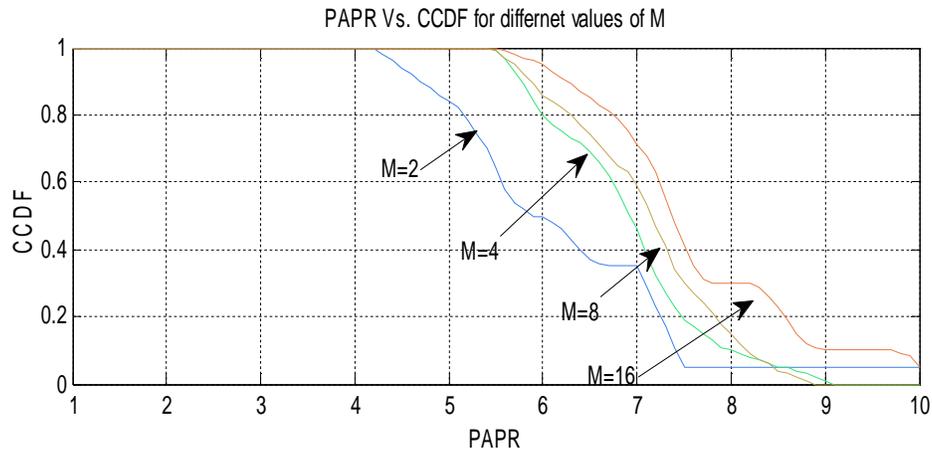

Figure 2. PAPR Vs CCDF for different M

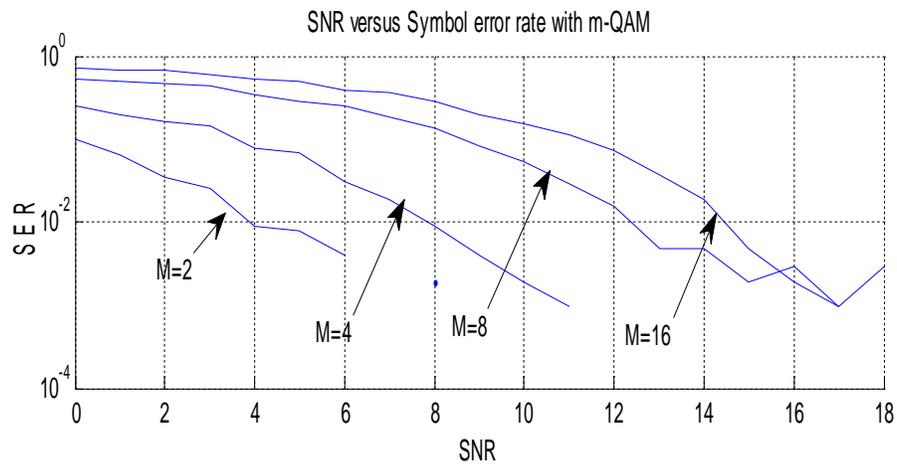

Figure 3. SNR Vs SER with different values of M

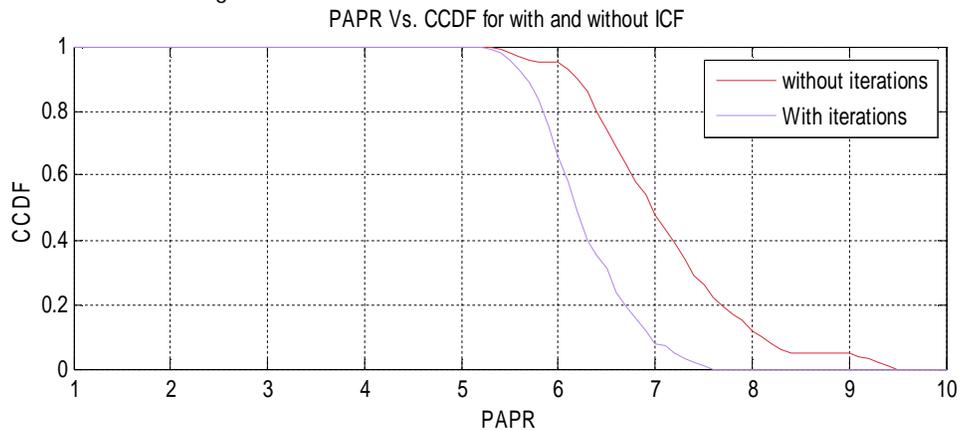

Figure 4. PAPR Vs CCDF with and without iterations

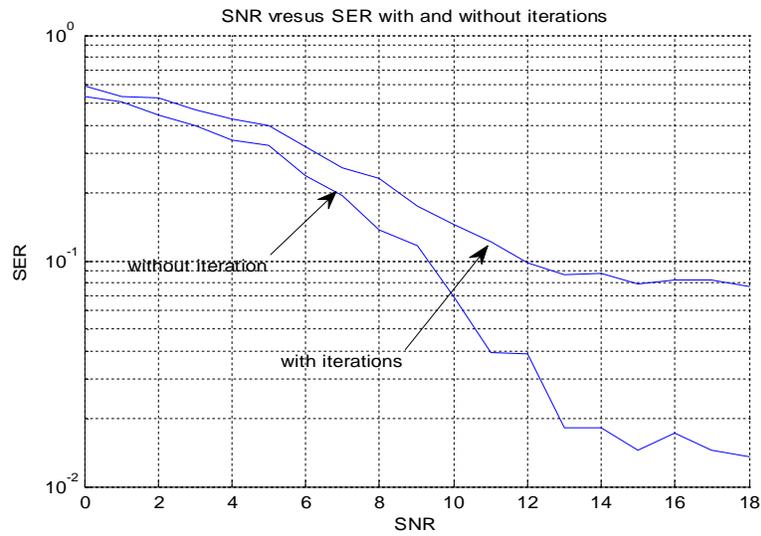

Figure 5. SNR Vs Error rate with and without iterations

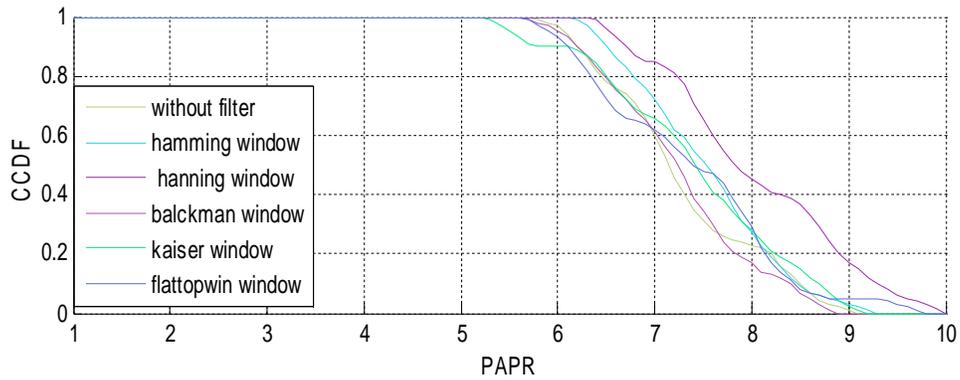

Figure 6. PAPR for different windows at M=8

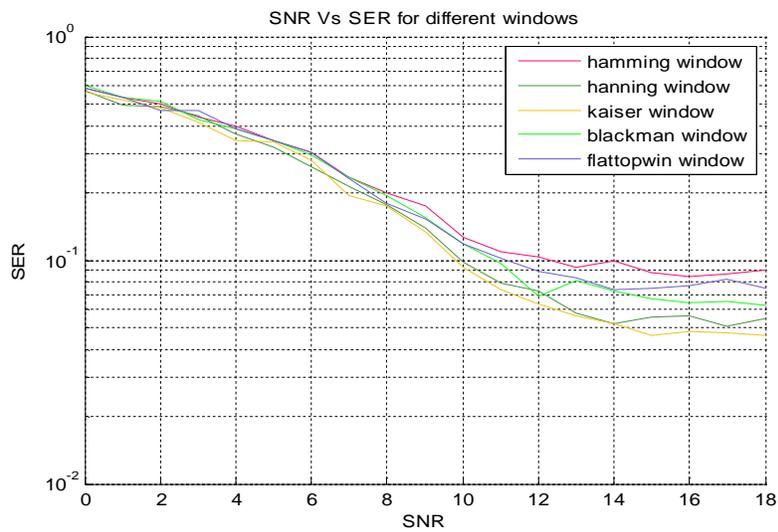

Figure 7. SNR Vs SER for various windows

V. RESULT

[1] As number of bits per symbol increases PAPR and Symbol error rate also increases as shown in figure 2 and 3.
[2] As the number of iterations increases PAPR reduces as shown in figure 4.

   PAPR=7.0607(without iteration)
   PAPR=6.2607(with iterations)

   But the symbol error rate increases as shown in figure 5.
[3] Figure 6 shows the SER for the different windows for out-of-band reduction in additive white Gaussian noise (AWGN) channel and 8 - QAM modulation scheme. Results are shown for varying filters. Got minimum PAPR for Kaiser window.

| WINDOW | PAPR |
|---|---|
| Kaiser | 7.3486 |
| Blackman | 7.3175 |
| Hanning | 7.2177 |
| Hamming | 6.9626 |
| Flattopwin | 6.8552 |

CONCLUSION:

It is shown that the PAPR can be significantly reduced with no increase in out-of-band power and little degradation in symbol error rate. The effect of different windows on the system SER is also studied. Results shows that due to fading the clipping noise will have less impact on system's SER. Further system can be optimized according to the above results.

REFERENCES:

[1] X. Li. And L. J. Cimini, " Effects of Clipping and Filtering on the Performance of OFDM," *IEEE Communications Letters*, Vol. 2; no 5, pp 131–133, May 1998.

[2] "Effect of Coding Schemes with Clipping and Filtering Method to Reduce PAPR on the Performance of OFDM System over Noisy Channels"Asian Journal of information Technology,Vol.10,2011

[3] "Modified Clipping and filtering technique for PAPR reduction of OFDM signals Used in WLAN",P.K.sharma ,IJEST Vol.2(10),2010,5337-5343.

[4] M. Pauli and H. P. Kuchenbecher, "On the reduction of the out of band radiation of OFDM signals", In Proceedings of IEEE International Conference on Communications, **(1998)**, pp. 1304-1308.

[5] "An Overview Of Techniques For Reducing Peakto Average Power Ratio And Its Selection Criteria For Orthogonal Frequency Division multiplexing Radio Systems"V. Vijayarangan, Dr. (Mrs) R. Sukanesh *Journal Of Theoretical And Applied Information Technology,Year 2009,Vol-5, No-5, E- Issn-* 1817-3195/Issn-1992-8645

[6] "Multi-carrier PAP reduction method using sub-optimal PTS with threshold," Oh-Ju Kwon and Yeong-Ho Ha, *IEEE Transactions on Broadcasting*, June. 2003, vol. 49, no. 2, PP. 232-53

.